\documentclass[fleqn,usenatbib]{mnras}

\usepackage{newtxtext,newtxmath}

\usepackage[T1]{fontenc}
\usepackage{ae,aecompl}
\DeclareRobustCommand{\VAN}[3]{#2}
\let\VANthebibliography\thebibliography
\def\thebibliography{\DeclareRobustCommand{\VAN}[3]{##3}\VANthebibliography}

\usepackage{graphicx}	
\usepackage{amsmath}	
\usepackage[utf8]{inputenc}
\usepackage{subfigure}

\title[The nature of the X-ray sources]{The nature of the X-ray sources in dwarf galaxies in nearby clusters from the KIWICS}

\author[\c{S}. \c{S}en et al.]{
\c{S}eyda \c{S}en,$^{1}$\thanks{E-mail: seyda.aydemir@sabanciuniv.edu}
Ersin G\"o\u{g}\"u\c{s},$^{1}$
Reynier F. Peletier,$^{2}$
Nelvy Choque-Challapa$^{3}$
and Amirnezam Amiri$^{4}$ $^{5}$ $^{6}$ $^{7}$ $^{8}$
\\
\\
$^{1}$Faculty of Engineering and Natural Sciences, Sabancı University, İstanbul 34956, Turkey\\
$^{2}$Kapteyn Astronomical Institute, University of Groningen, P. O. Box 800, 9700 AV Groningen, The Netherlands\\
$^{3}$Instituto de Astronom\'{i}a y Ciencias Planetarias, Universidad de Atacama, Copayapu 485, Copiap\'{o}, Chile\\
$^{4}$Dipartimento di Fisica e Astronomia, Universit\`{a} di Firenze, Via G. Sansone 1, 50019, Sesto Fiorentino (Firenze), Italy\\
$^{5}$Iranian National Observatory, Institute for Research in Fundamental Sciences (IPM), Tehran, Iran\\
$^{6}$Departamento de Astrof\'{i}sica, Universidad de La Laguna, E-38200, La Laguna, Tenerife, Spain \\
$^{7}$Instituto de Astrof\'{i}sica de Canarias E-38205, La Laguna, Tenerife, Spain\\
$^{8}$Department of Physics, University of Arkansas, 226 Physics Building, 825 West Dickson Street, Fayetteville, AR 72701, USA\\
}

\date{Accepted XXX. Received YYY; in original form ZZZ}

\pubyear{2023}

\begin{document}
\label{firstpage}
\pagerange{\pageref{firstpage}--\pageref{lastpage}}
\maketitle

\begin{abstract}
We present a deep search for and analysis of X-ray sources in a sample of dwarf galaxies (M$_{r}$ < -15.5 mag) located within twelve galaxy clusters from the Kapteyn IAC WEAVE INT Cluster Survey (KIWICS) of photometric observations in the $\textit{r}$ and $\textit{g}$ using the Wide Field Camera (WFC) at the 2.5-m Isaac Newton telescope (INT). We first investigated the optical data, identified 2720 dwarf galaxies in all fields and determined their characteristics; namely, their colors, effective radii, and stellar masses. We then searched the $\textit{Chandra}$ data archive for X-ray counterparts of optically detected dwarf galaxies. We found a total of 20 X-ray emitting dwarf galaxies, with X-ray flux ranging from 1.7$\times10^{-15}$ to 4.1$\times10^{-14}$ erg cm$^{-2}$ s$^{-1}$ and X-ray luminosities varying from 2$\times10^{39}$ to 5.4$\times10^{41}$ erg s$^{-1}$. Our results indicate that the X-ray luminosity of the sources in our sample is larger than the Eddington luminosity limit for a typical neutron star, even at the lowest observed levels. This leads us to conclude that the sources emitting X-rays in our sample are likely black holes. Additionally, we have employed a scaling relation between black hole and stellar mass to estimate the masses of the black holes in our sample, and have determined a range of black hole masses from 4.6$\times10^{4}$ to 1.5$\times10^{6}$ M$\sun$. Finally, we find a trend between X-ray to optical flux ratio and X-ray flux. We discuss the implications of our findings and highlight the importance of X-ray observations in studying the properties of dwarf galaxies.

\end{abstract}




\section{Introduction}

Galaxy clusters are the largest and most massive gravitationally-bound structures in the Universe. According to current cosmological theories of large scale formation, galaxy clusters have grown hierarchically by the merging of smaller virialized haloes \citep{press.1974,white.rees.1978,blumenthal.1984}. They allow us to study numerous astrophysical phenomena. In particular, the investigations of the formation of early-type galaxies whose fraction is much higher in clusters as determined by the morphology-density relation \citep{dressler80}. This relation also reveals that the fraction of dwarf galaxies is even higher.

Dwarf galaxies constitute the most numerous subset of galaxies in the Universe  \citep{binggeli.88}. However, the observed number of dwarfs is lower by about two orders of magnitude than expected from the current models of galaxy formation \citep{Moore99}. This is the so-called missing satellite problem.  Moreover, the processes driving the formation of dwarf galaxies and how the environment affects their evolution are still poorly understood. 

The problem of the dwarf invisibility is generally attributed to feedback processes which ejects most of the baryons, thus, making them difficult or impossible to be detected. \citet{silkandmamon12} review that three fundamental mechanisms for dwarf galaxy feedback are reionization of the Universe at early epochs, supernovae (SNe) and (ram pressure and tidal) stripping. Alternatively, active galactic nuclei (AGN) driven outflows from black holes (BHs) could contribute to the feedback mechanism. This scenario has been gaining support through in depth studies in recent years \citep{silkandnusser.10}. However, none of these models has so far has provided clear a solution. More accurate census of dwarf galaxies and revealing their properties will allow us to better understand their formation mechanisms and can be a benchmark for our cosmological models.

In massive galaxies, AGN feedback mechanism has been added as a regular component of the galaxy formation model since it seems that every massive galaxy has a supermassive black hole (SMBH) in its centre \citep{kormendyandrichstone95}. In dwarf galaxies, on the other hand, the presence of massive BHs (mBH; $M_{BH}$ = 10$^{4}$ $\sim$ 10$^{6}$ $M_{\sun}$; \citealp{greene.2020,mezcua.2017}) is becoming increasingly evident, thanks to studies of individual galaxies to large scale surveys (e.g. \citealp{reines.2011,reines.2013,moran.2014,mezcua.2020,birchall.2020}). However, the occupation fraction of BHs in dwarf galaxies is still debated. The best evidence for BHs in dwarf galaxies comes from X-ray \citep{Kormendy.2013,pardo.2016} besides optical emission line studies \citep{baldassare.2016}. These investigations indicate that AGNs are found in roughly one percent of dwarf galaxies.

X-ray observations are ideal for the detection of galaxy clusters since the X-ray emission is proportional to the square of the gas density, making it easier to identify clusters with cool cores. In addition,  the projection effects can be completely overcome in the X-ray band. The first cluster samples were compiled from the first all-sky X-ray survey with \textit{Uhuru} and subsequent further observations were found more objects with \textit{HEAO-1}, \textit{Ariel-V}, \textit{Einstein}, \textit{EXOSAT} and a centerpiece \textit{ROSAT}. Later deep pointed observations with the current generation of X-ray satellites \textit{XMM-Newton}, \textit{Chandra}\footnote{To see a list of recent and ongoing Chandra and XMM Newton surveys http://cxc.harvard.edu/xraysurveys/surveys.html} and \textit{Suzaku} have remarkably changed our X-ray view of clusters and their galaxies to investigate their evolutionary properties (see \citealt{Rosati.2002}, for a review).

The resolving power of \textit{Chandra} and the low background of its ACIS instrument provide us an ideal combination to detect faint sources in nearby galaxies. Several studies have taken this advantages to study the fraction of AGNs in nearby late-type galaxies and have successfully detected X-ray nuclei in star-forming galaxies \citep{ghosh.2008,grier.2011,she.2017a}, late-type bulgeless spirals \citep{she.2017b} and dwarf irregular galaxies \citep{lemons.2015}. For the local AGNs selected from the Sloan Digital Sky Survey (SDSS), more attention in X-rays has been devoted to perform detailed spectral and timing analysis using sufficiently deep observations (e.g. \citealt{moran.2005,greene.2007,dong.2012,jin.2016}). In addition, based on mainly archival observations, detailed characterizations of X-ray nuclei in nearby, lower mass early-type galaxies in the Virgo cluster \citep{ghosh.2008,gallo.2010}, and the Fornax cluster \citep{lee.2019} were performed.

In this paper, we performed one of the first investigations to unveil X-ray sources within dwarf galaxies in nearby galaxy clusters. Our search for X-ray emission using deep archival Chandra observations focused on dwarf galaxies up to a redshift of z $\sim$0.03. These dwarf galaxies were identified with optical data in order to bring some light into galaxy evolution and transformation processes. It is clear that studying dwarf galaxies in cluster environments of different properties in both optical and X-rays would yield deeper understanding about their evolution.

Here, we use a sample of twelve nearby clusters from Kapteyn IAC WEAVE INT Cluster Survey (KIWICS) and present the results of our comprehensive investigations to uncover X-ray emission from dwarf galaxies in these galaxy clusters. In the next section, we describe optical and X-ray observations and explain our methodology to reduce them. In Section 3, we present the main results. We conclude, in section 4, with a discussion of these results and a comparison of X-rays versus optical data. 

Throughout this paper, we assume a $\Lambda$CDM cosmology with $\Omega_{m}$ = 0.3, $\Omega_{\lambda}$ = 0.7, and H$_{0}$ = 70 kms$^{-1}$Mpc$^{-1}$.

\section{Observations and Data Reduction}
\subsection{Optical Observations}
\label{subsec:opticalobservation}
Our optical observational data come from a deep photometric survey of galaxy clusters: The Kapteyn IAC WEAVE INT Cluster Survey (KIWICS, PIs R. Peletier and J.A. Lopez-Aguerri). This survey consisted in imaging of 47 X-ray selected, nearby (0.02 < $\textit{z}$ < 0.04) galaxy cluster \citep{piffaretti.2011} in the Northern hemisphere, and will be ideal for studying dwarfs and low-surface-brightness galaxies (LSB). All observational data were obtained by using the Wide Field Camera (WFC) at the 2.5-m Isaac Newton Telescope (INT)
in La Palma, Spain. The full KIWICS sample was selected as a preparation for the future spectroscopic WEAVE Cluster Survey to be carried out with WHT Enhanced Area Velocity Explores (WEAVE) spectrograph \citep{dalton.2016}. We use the two broadband Sloan filters \textit{g} and \textit{r} with total integration times of $\sim$ 1800 s and $\sim$ 5400 s, respectively. The observations cover each cluster up to at least 1 $R_\textit{200}$ with a dithering pattern of individual exposures of 210s. A comprehensive description of the observational strategy and data reduction processes can be found in \cite{pavelmancera.2018,pavelmancera.2019} and \cite{nelvy.2021}. Here, we briefly summarise the main aspects.

The data reduction was done by using the Astro-WISE \citep{McFarland.2013} environment following the same routine as explained in there. The data reduction was performed in two main steps; the first step contains the standard instrumental corrections, namely, applying bias subtraction and flat-fielding corrections. At this stage, weight maps were also generated for each frame. These weight maps contain information about bad pixels or saturated pixels (from hot and cold pixel maps), as well as the expected noise associated with each pixel and cosmic rays. The second step deals with the sky subtraction, after which astrometric and photometric corrections were applied. For this purpose, a set of standard stars were observed during each night of the observations with the SDSS DR14 catalogue \citep{abolfathi.2018}. Finally, all the cluster frames corrected for bad pixels and cosmic rays. The astrometric solutions were computed by making use of the publicly available software SCAMP \citep{bertin.2006}. The astrometry of our final mosaic has an rms of $\sim$ 0.2$''$. They are median-stacked to produce a deep coadded mosaic with re-sampled to a scale of 0.2$''$ per pixel. 

\begin{figure}
	\includegraphics[width=1.0\columnwidth]{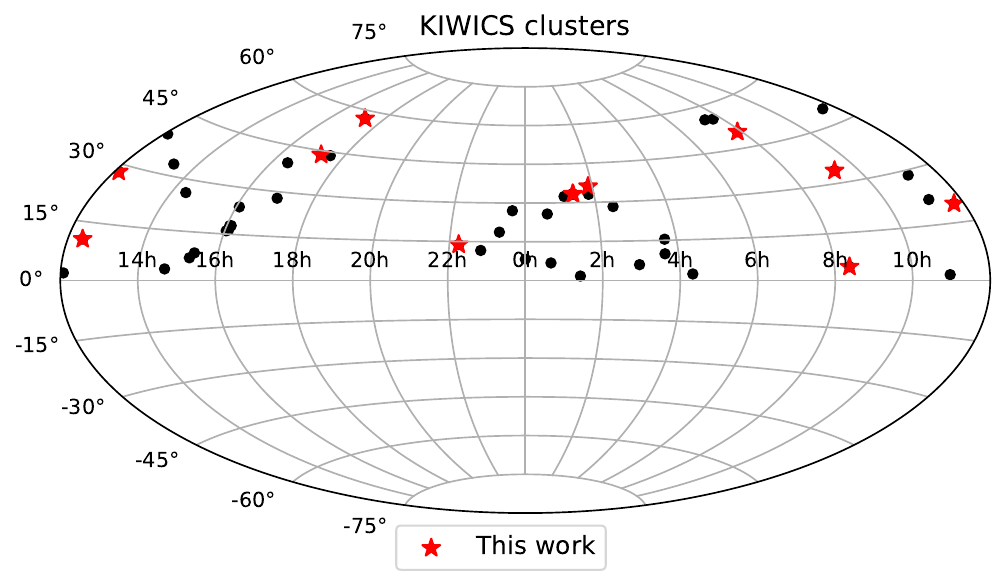}
    \caption{Sky map of all of the clusters being surveyed in KIWICS and those studied in this work represent in red star symbols.}
    \label{fig:skymap}
\end{figure}

The galaxy clusters studied here come from the sample of \cite{nelvy.2021} that have a seeing < 1.6$''$ FWHM with redshift lower than 0.03. In this study, we analyse dwarf galaxies in these clusters by combining optical and X-ray data. To classify objects as dwarf galaxies, we employed a criterion of M$_{r}$ > $-$19.0 mag. This threshold adheres to the established convention, as defined by  \citet{binggeli.88}, which designates dwarf galaxies as those with M$_{B}$ > $-$18.0 mag and assumes a color index $\textit{B - r}$ $\sim$ 1.0 mag for such galaxies. Additionally, we implemented a lower luminosity cutoff at M$_{r}$ = $-$15.5 mag. Galaxies fainter than this limit were excluded from our analysis to reduce potential contamination from background objects.

\begin{table*}
 \caption{Properties of our sample of galaxy clusters. }
 \label{tab:properties}
 \begin{tabular}{ccccccccc} 
	    \hline
		Name &  RA  & Dec & Redshift &  $\sigma_\text{e}$ & Seeing $\textit{r}$-band  & Seeing $\textit{g}$-band & $M_{500}$ & $R_{500}$ \\
        & (deg) & (deg)& z &(km s$^{-1}$)& (") & (")& 10$^{14}$ $M_{\sun}$ & Mpc \\
        (1) & (2) & (3) & (4) & (5) & (6) & (7) & (8) & (9) \\
        \hline       
A1367	&	176.152	&	19.759	&	0.0214	&	581	$\pm$	64		&	1.5	&	1.6	&	2.14	&	0.90	\\
A262	&	28.188	&	36.157	&	0.0163	&	402	$\pm$	24		&	1.5	&	1.4	&	1.19	&	0.74	\\
RXJ0123.2+3327	&	20.801	&	33.461	&	0.0146	&	483	$\pm$	95		&	1.6	&	1.3	&	0.36	&	0.50	\\
RXJ0123.6+3315	&	20.921	&	33.261	&	0.0164	&	483	$\pm$	95		&	1.6	&	1.3	&	0.61	&	0.60	\\
RXCJ0751.3+5012	&	117.844	&	50.213	&	0.0228	&	360	$\pm$	24		&	1.3	&	1.3	&	0.42	&	0.52	\\
RXCJ1206.6+2811	&	181.656	&	28.184	&	0.0283	&	381	$\pm$	42		&	1.5	&	1.4	&	0.42	&	0.52	\\
RXCJ1223.1+1037	&	185.777	&	10.624	&	0.0258	&	302	$\pm$	12		&	1.6	&	1.5	&	0.56	&	0.58	\\
RXCJ1715.3+5724	&	258.841	&	57.408	&	0.0276	&	475	$\pm$	29		&	1.3	&	1.5	&	0.87	&	0.67	\\
RXCJ0919.8+3345	&	139.955	&	33.760	&	0.0230	&	299	$\pm$	19		&	1.4	&	1.4	&	0.26	&	0.45	\\
RXCJ1714.3+4341	&	258.578	&	43.690	&	0.0276	&	176	$\pm$	22		&	1.3	&	1.5	&	0.31	&	0.48	\\
RXCJ2214.8+1350	&	333.720	&	13.847	&	0.0253	&	351	$\pm$	07		&	1.3	&	1.3	&	0.32	&	0.48	\\
ZwCL1665	&	125.798	&	04.356	&	0.0293	&	382	$\pm$	15		&	1.3	&	1.4	&	0.73	&	0.63	\\
	\hline
  \end{tabular}
  \vspace{1ex}

{\raggedright \textbf{Note.} (1) galaxy cluster name; (2) and (3) right ascension and declination in J 2000; (4) and (5) redshift and velocity dispersion from \citet{nelvy.2021}; (6) and (7) mean seeing during the observation in the $\textit{r}$ and $\textit{g}$ band; (8) and (9) mass ($M_{500}$) and radius ($R_{500}$) from \citet{piffaretti.2011}. \par}

 \end{table*}

\subsubsection{Identification of dwarf galaxies }
We use SExtractor \citep{bertin.1996,holwerda.2005} based on the stellarity CLASS$_\text{--}$STAR and FLAG parameters to detect potential dwarf candidates. The CLASS$_\text{--}$STAR parameter ranges from 0 to 1; objects close to 0 are likely to be extended objects and close to 1 are more likely to be point sources. Galaxies are defined as object with CLASS$_\text{--}$STAR $\leq$ 0.2 in both \textit{g} and \textit{r} filters, and with FLAGS$_\text{r}$ = 0. This criterion is used to exclude objects that are either blended with other nearby objects or have poor quality photometry.

\begin{figure*}
    \subfigure{\includegraphics[width=0.49\textwidth]{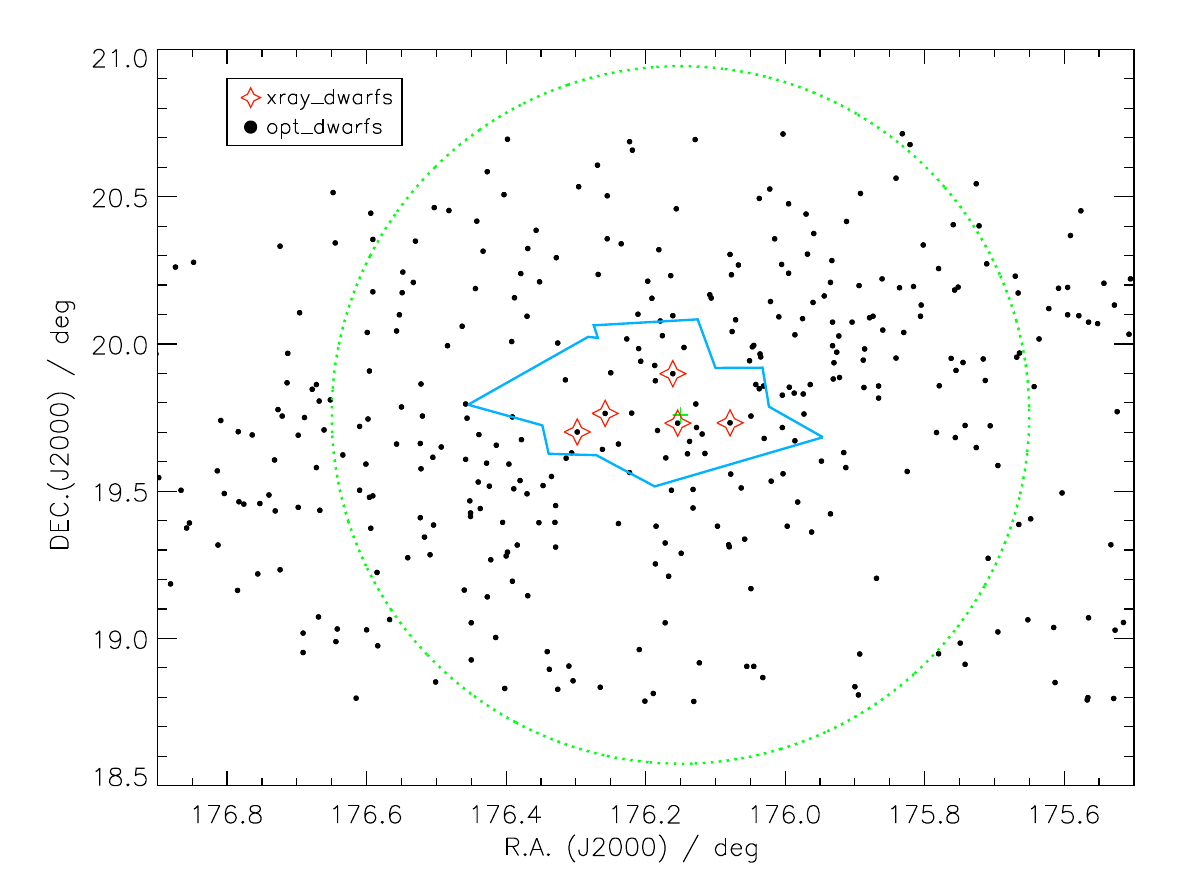}}
    \subfigure{\includegraphics[width=0.49\textwidth]{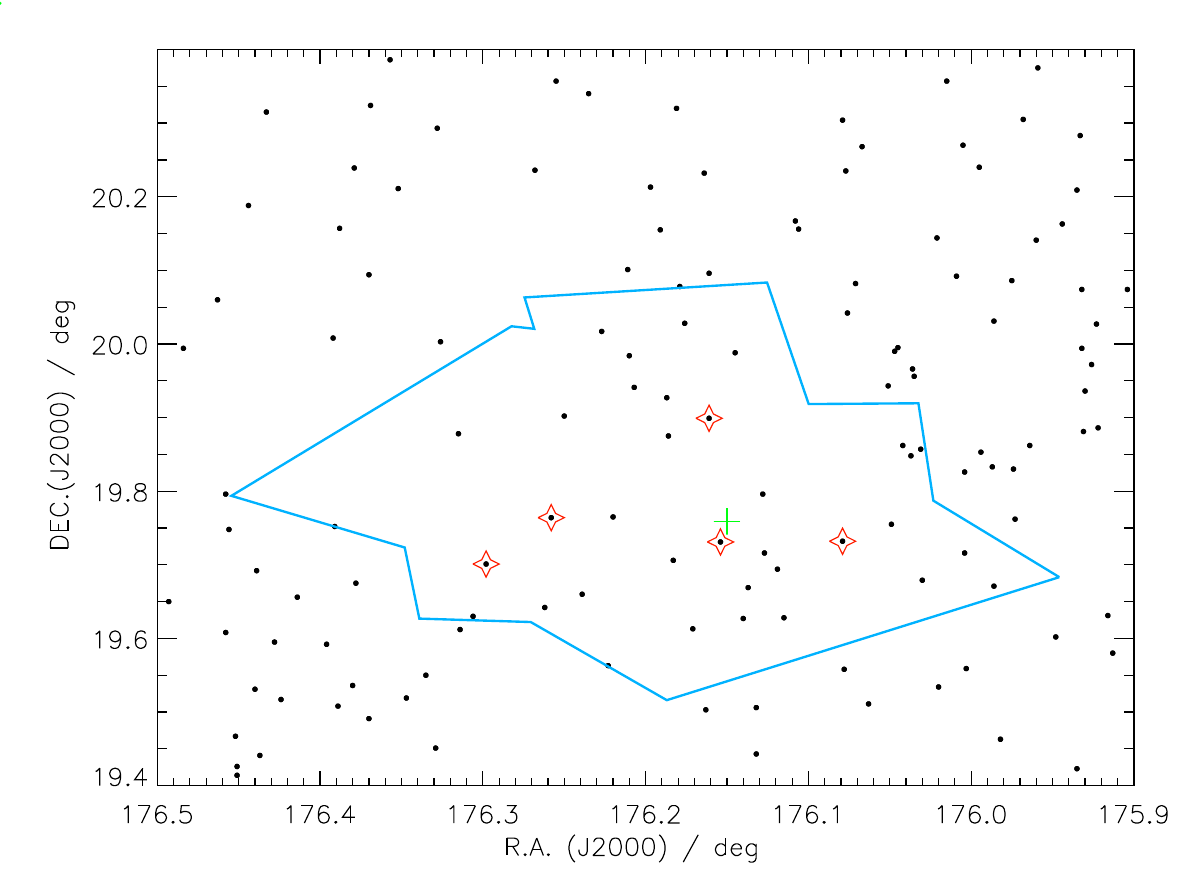}}
    \caption{(Left) Map of the A1367 cluster. The black symbols correspond to dwarf galaxies identified in optical. The red stars represent matching galaxies with X-rays. The green cross shows the X-ray center from Ebeling et al. (1998), the green dotted circle indicates the estimated R$_{200}$ radius, and the blue area indicates the combined field covered by multiple Chandra observations. (Right) Zoomed in view of the X-ray centre of the galaxy cluster.}
    \label{fig:map_a1367}
\end{figure*}

To effectively eliminate background objects from the SExtractor catalogue, we firstly performed a cut based on the color \textit{g - r} $\leq$ 1.0 mag to exclude all galaxies redder than this limit. This limit corresponds to a 12 Gyr old stellar population with supersolar metallicity, respectively \citep{worthey1994}. As a final step of our selection of the dwarf candidates, a visual cleaning was done in order to remove artefacts if present. We exclude all the objects if they are background, artefacts, interaction, or fainter than \textit{m$_\text{r}$} = 20 mag, as their visual selection become inaccurate (see Section 3 in \citealp{nelvy.2021}, for details). In the end, we find that there are overall 2720 dwarf galaxies detected in all of these 12 clusters of the KIWICS survey fields. As an example, we present the dwarf galaxies in A1367 in Figure \ref{fig:map_a1367}.

In order to obtain more accurate photometric measurements for the identified galaxies, we used GALFIT \citep{peng.2010} via a fit to their light profiles. The process was carried out according to the methodology outlined in \cite{venhola.2018} for all probable cluster members, excluding objects with an SExtractor ISORAREA-IMAGE value of less than 200 pixels. This exclusion was primarily due to the fact that smaller objects tend to result in poor fits due to their faintness (apparent \textit{r}-band magnitudes fainter than 22 mag). We used the central coordinates, isophotal magnitudes, and semi-major axis lengths (all obtained using SExtractor) as inputs for the photometric investigation. We adopt the non-circularized effective radius, which represents the length of the semimajor axis of an ellipse that best fits the isophotes enclosing half of the total light emitted by a galaxy.
A single Sérsic function was applied to each object in both the \textit{r} and \textit{g} bands. However, due to the limited resolution of our images, it was not possible to discern a nucleus in the galaxies. For the objects that have not a clear nucleus, the center coordinates are kept fixed when determining the SExtractor magnitudes. 

To provide a comprehensive analysis, we also estimate the stellar masses of the sample galaxies. To accomplish this, we use the relation observed between the (g$-$r) color and mass to-light (M/L) ratio by \citet{roediger.2015}:

    \begin{equation}
        \log(M_{*}/L_{r}) = 1.629 \times (g-r) - 0.792 
    \end{equation}

Our all dwarf samples and X-ray emitting dwarf samples have a median (mean) stellar mass of 4.93 (5.33) $\times10^{8}$ $M_{\sun}$ and 7.16 (10.90) $\times10^{8}$ $M_{\sun}$, respectively.

\subsection{X-ray Observations}
We searched the Chandra data archive for X-ray observations covering the fields of our sample of 12 galaxy clusters with minimum exposure times of 10 ks in order to ensure significant detection. We find that one cluster (RXCJ1714.3+4341) was not observed with Chandra. The exposure times of X-ray observations of two clusters, namely RXCJ0919.8+3345 and ZwCL1665 were 3 and 2 ks, respectively. Therefore, these two observations were not used. The remaining nine clusters were found to have deep enough Chandra observations in one or multiple pointings. We list the observation IDs of these data sets in Table~\ref{tab:result_xray}.

The Chandra data were reduced using the Chandra Interactive Analysis of Observations (\textsc{ciao}) software version 4.13 with \textsc{CALDB} version 4.9.5. As the focal point instruments, we employed data recorded with a back-illuminated ACIS-S chip (S3) or the four front-illuminated chips of ACIS-I (I0-I3). Note that the other chips on ACIS-S were generally not useful for our purposes due to the fact that the point spread function (PSF) becomes large at large off-axis angles.

To classify optical counterparts to our Chandra X-ray data, we place both the Chandra and KIWICS data onto the International Celestial Reference System (ICRS) by finding matches between stars or background galaxies in the Two Micron All-Sky Survey Point Source Catalog (2MASS-PSC, \citealp{skrutskie.2006}). We have already performed astrometry for KIWICS data (see Section~\ref{subsec:opticalobservation}). In order to improve the astrometry of Chandra/ACIS images relative to the KIWICS images, we use our own pipeline which aligns the images using the astroalign Python package \citep{beroiz.2020} and the CIAO task \textsc{wcs-match} script. Using the transformation matrices obtained from the analysis, the aspect solution files and the coordinate parameters are updated in all of the X-ray event files. We set an upper bound for the rms residual between optical and X-ray positions as $\sim$ 0.02$''$.

We identified X-ray sources within our aligned Chandra images using the \textsc{wavdetect} tool of CIAO. It is a Mexican-Hat wavelet-based source detection algorithm on the full energy band (0.5$-$10 keV) with the false detection threshold value set to 10$^{-6}$. The wavelet scales are set to 1.0, 2.0, 4.0 and 8.0 pixels. We detected 487 X-ray sources in all nine fields that we investigated. Finally, we obtained the matching pairs of the coordinates of Chandra X-ray sources and optical dwarf source lists by allowing the two to be at most 0.5$''$ apart from each other. In total, 20 dwarf galaxy sources were detected in X-rays.  We list these 20 sources in Table \ref{tab:obs_xray}. We also illustrate the optical and X-ray paired galaxies in A1367 in Figure \ref{fig:map_a1367}.

\begin{table}
 \caption{Details of Chandra X-Ray Observations for the 20 X-ray Emitting Galaxies}
 \label{tab:obs_xray}
 \begin{tabular}{lcccc} 
	    \hline 
	    Name	 & ObsID  & Start Date and Time  & Exposure  & Count Rate$^{(1)}$\\
	    & & (UTC)   & (ks) & (10$^{-4}$ cts s$^{-1}$)\\
	    \hline 
A1367-1	&	4189	&	2003-01-24 10:29:13	&	48	 & 25.2 $\pm 3.0$\\
& 17199 & 2015-01-30 21:05:47 & 38 & \\
& 17201 & 2016-01-31 13:37:06 & 61 &\\
A1367-2	&	514	&	2000-02-26 10:44:03	&	41	& 	18.6	$\pm	3.6	$\\
& 17199 & 2015-01-30 21:05:47 &38& \\
& 17200 & 2015-11-05 00:56:42 & 40&\\
& 17201 & 2016-01-31 13:37:06 & 61&\\
A1367-3	&	17199	&	2015-01-30 21:05:47	&	38 &	6.4	$\pm	2.5	$	\\
& 17200 & 2015-11-05 00:56:42 & 40&	\\
A1367-4	&	17199	&	2015-01-30 21:05:47	&	38 & 6.2	$\pm	2.4	$	\\
& 17200 & 2015-11-05 00:56:42 & 40&\\
& 17201 & 2016-01-31 13:37:06 & 61&\\
A1367-5	&	17199	&	2015-01-30 21:05:47	&	38	&	10.8	$\pm	3.0	$\\
& 17200 & 2015-11-05 00:56:42 & 40&\\
& 17201 & 2016-01-31 13:37:06 & 61& \\
A262-1	&	7921	&	2006-11-20 03:35:12	&	111 &	54.4	$\pm	3.6	$\\
A262-2	&	7921	&	2006-11-20 03:35:12	&	111	&	8.8	$\pm	2.0	$\\
A262-3	&	7921	&	2006-11-20 03:35:12	&	111	&	7.4	$\pm	3.1	$\\
RXJ0123-1	&	2882	&	2002-01-08 04:47:16	&	44	&	42.6	$\pm	5.0	$\\
RXCJ0751-1	&	15170	&	2013-05-14 07:59:03	&	98	&	7.6	$\pm	1.2	$\\
RXCJ0751-2	&	15170	&	2013-05-14 07:59:03	&	98	&	2.9	$\pm	1.3	$\\
RXCJ0751-3	&	15170	&	2013-05-14 07:59:03	&	98	&	2.3	$\pm	0.9	$\\
RXCJ1206-1	&	6939	&	2006-02-16 02:40:12	&	36	&	20.6	$\pm	3.7	$\\
RXCJ1223-1	&	3232	&	2003-02-04 16:09:47	&	30	&	15.6	$\pm	3.6	$\\
RXCJ1715-1	&	4194	&	2003-09-17 08:33:38	&	47	&	2.5	$\pm	2.4	$\\
RXCJ1715-2	&	4194	&	2003-09-17 08:33:38	&	47	&	3.8	$\pm	2.0	$\\
RXCJ1715-3	&	4194	&	2003-09-17 08:33:38	&	47	&	13.1	$\pm	2.3	$\\
RXCJ2214-1	&	6392	&	2006-01-12 22:33:07	&	33	&	6.1	$\pm	3.0	$\\
RXCJ2214-2	&	6392	&	2006-01-12 22:33:07	&	33	&	8.3	$\pm	2.4	$\\
RXCJ2214-3	&	6392	&	2006-01-12 22:33:07	&	33	&	9.8	$\pm	2.2	$\\

\hline
\end{tabular}
{\raggedright \textbf{Note.} (1) Background subtracted count rates in the 0.5$-$10 keV band}.

\end{table}

\begin{table*}
 \caption{Results of X-ray Spatial and Spectral Investigations}
 \label{tab:result_xray}
 \begin{tabular}{lccclccc} 
	    \hline 
	 Name	 & RA  & Dec  & $N_{H}$ & Photon Index &  Flux & $log L_{x}$ & $\chi^{2}$ / dof  \\
	  & (deg) & (deg) & (10$^{20}$ cm$^{-2}$) & ~~~~$\Gamma$ &(10$^{-14}$ erg cm$^{-2}$ s$^{-1}$)& (erg s$^{-1}$)  \\
        \hline
A1367-1	&	176.152	&	19.893	&	1.80	&	$1.77^{+0.33}_{-0.30}$	&    4.05 $\pm	0.25	$	&	40.62	&	26.45/36	\\
A1367-2	&	176.160	&	19.735	&	1.80	&	$1.72^{+0.29}_{-0.26}$	&	2.07$\pm	0.18	$	&	40.33	&	65.18/69	\\
A1367-3	&	176.257	&	19.764	&	1.80	&	1.75		            &	1.42 $\pm	0.34	$	&	40.17	&	4.85/5	\\
A1367-4	&	176.298	&	19.702	&	1.80	&	1.75			        &	1.19$\pm	0.26	$	&	40.09	&	16.38/13	\\
A1367-5	&	176.070	&	19.738	&	1.80	&	1.75		            &	2.37$\pm	0.27$	&	40.39	&	26.79/26	\\
A262-1	&	28.210	&	36.155	&	6.80	&	$2.17^{+0.21}_{-0.20}$	&	2.94$\pm	0.18	$	&	40.25	&	128.19/72	\\
A262-2	&	28.170	&	36.198	&	6.80	&	2.00		            &	0.56 $\pm	0.10	$	&	39.53	&	3.82/5	\\
A262-3	&	28.261	&	36.160	&	6.80	&	$1.92^{+1.65}_{-0.98}$	&	0.67$\pm	0.14	$	&	39.60	&	43.86/50	\\
RXJ0123-1	&	20.892	&	33.252	&	5.26	&	$2.39^{+0.49}_{-0.40}$ &	2.74$\pm	0.30	$	&	40.12	&	35.22/24	\\
RXCJ0751-1	&	117.609	&	50.156	&	5.61	&	$0.96^{+0.66}_{-0.60}$ &	1.66$\pm	0.21	$	&	40.29	&	6.54/7	\\
RXCJ0751-2	&	117.864	&	50.169	&	5.61	&	$1.54^{+1.45}_{-1.11}$	&	0.59$\pm	0.16	$	&	39.84	&	2.71/6	\\
RXCJ0751-3	&	117.671	&	50.175	&	5.61	&	1.50		&	0.17$\pm	0.03	$	&	39.30	&	5.79/3	\\
RXCJ1206-1	&	181.712	&	28.108	&	1.72	&	$1.02^{+0.67}_{-0.58}$	&	2.93$\pm	0.25	$	&	40.73	&	10.85/9	\\
RXCJ1223-1	&	185.742	&	10.630	&	2.71	&	$2.49^{+1.20}_{-1.00}$	&	0.85$\pm	0.14	$	&	40.11	&	1.98/5	\\
RXCJ1715-1	&	258.892	&	57.450	&	2.12	&	1.61	                &	0.29$\pm	0.04	$	&	39.71	&	3.28/5	\\
RXCJ1715-2	&	258.768	&	57.447	&	2.12	&	1.61	                &	0.41$\pm	0.11	$	&	39.85	&	5.29/3	\\
RXCJ1715-3	&	258.818	&	57.337	&	2.12	&	$1.61^{+0.65}_{-0.56}$	&	1.66$\pm	0.28	$	&	40.46	&	1.74/5	\\
RXCJ2214-1	&	333.788	&	13.972	&	4.31	&	1.50	                &	0.93$\pm	0.32	$	&	40.13	&	7.17/9	\\
RXCJ2214-2	&	333.640	&	13.810	&	4.31	&	$1.26^{+1.18}_{-1.00}$	&	1.33$\pm	0.25$	&	40.29	&	7.21/5	\\
RXCJ2214-3	&	333.815	&	13.915	&	4.31	&	$1.58^{+0.91}_{-0.79}$	&	1.55$\pm	0.29	$	&	40.35	&	7.29/6	\\

		\hline
  \end{tabular}
 \end{table*}

For each of these X-ray emitting systems, we performed X-ray spectral analysis to determine their X-ray flux as follows: We extracted source spectra for 18 sources from a circular region of 10$''$ radius centered at their X-ray positions listed in Table~\ref{tab:result_xray}. For two sources, in particular A262$-$2 and RXCJ2214$-$1, the radius was set to be 12$''$ due to their extended nature. We then grouped these source spectra in order for each spectral bin to contain 10 source counts. The background spectra were extracted from circular aperture of the same radii from nearby source-free regions. To take into account time-dependent and position-dependent ACIS responses, the corresponding response and ancillary response files are also extracted per observation. For the five sources in the A1367 galaxy cluster for which there were multiple Chandra observations, we modeled these multiple spectra for each source simultaneously by linking the power law index parameters so that the fit would yield a joint power law index for each source.

We fit each background subtracted spectrum with the power law model attenuated by the interstellar hydrogen population using XSPEC \citep{Arnaud96}. We fix the HI column density \textit{N$_H$} with corresponding value for each cluster that is the Galactic absorption towards these clusters\footnote{Obtained from NASA's HEASARC N$_H$ tool, https://heasarc.gsfc.nasa.gov/cgi-bin/Tools/w3nh/w3nh.pl}. We initially allowed the power law index to vary to perform the fits. We find that the spectra of 12 sources can be represented with the power law model whose indices range from about 1 to 2.5 (see Table~\ref{tab:result_xray}). However, the index parameter for the spectra of eight X-ray source cannot be constrained. In those cases, we fixed the power law index at the average value of constrained indices of the other cluster members. We finally used the best fit parameters to calculate the X-ray flux in the band 0.5$-$10 keV (see Table~\ref{tab:result_xray}). We list \textit{g $-$ r} color, M$_r$ magnitudes, effective radius as well as stellar mass (M$_{*}$) estimates of these 20 X-ray emitting dwarf galaxies in Table \ref{tab:result_optical}.

\begin{table}
 \caption{Optical Properties of the 20 X-ray Emitting Galaxies}
 \label{tab:result_optical}
 \begin{tabular}{lcccc} 
	    \hline 
	 Name	 &  \textit{g $-$ r}  & M$_{r}$ & $R_{e}$ & $\textit{log $M_*$}$ \\
	  & (mag) & (mag) & (kpc) & ($M_{\sun}$) \\
        \hline
A1367-1	    &	0.68	$\pm	0.04	$	&	-18.40	$\pm	0.09	$	&	1.02	$\pm	0.01	$	&	9.55	\\
A1367-2	    &	0.68	$\pm	0.03	$	&	-16.08	$\pm	0.22	$	&	0.88	$\pm	0.01	$	&	8.73	\\
A1367-3	    &	0.70	$\pm	0.03	$	&	-15.56	$\pm	0.25	$	&	2.39	$\pm	0.07	$	&	8.62	\\
A1367-4	    &	0.69	$\pm	0.03	$	&	-16.34	$\pm	0.13	$	&	1.31	$\pm	0.02	$	&	8.48	\\
A1367-5	    &	0.68	$\pm	0.06	$	&	-18.96	$\pm	0.14	$	&	0.71	$\pm	0.01	$	&	8.43	\\
A262-1	    &	0.76	$\pm	0.04	$	&	-18.65	$\pm	0.11	$	&	0.93	$\pm	0.00	$	&	9.78	\\
A262-2      &	0.55	$\pm	0.05	$	&	-15.93	$\pm	0.24	$	&	0.70	$\pm	0.01	$	&	8.83	\\
A262-3	    &	0.63	$\pm	0.03	$	&	-16.81	$\pm	0.16	$	&	1.11	$\pm	0.09	$	&	8.35	\\
RXJ0123-1	&	0.69	$\pm	0.04	$	&	-16.39	$\pm	0.20	$	&	0.85	$\pm	0.01	$	&	8.75	\\
RXCJ0751-1	&	0.79	$\pm	0.04	$	&	-16.95	$\pm	0.12	$	&	1.23	$\pm	0.01	$	&	9.35	\\
RXCJ0751-2	&	0.40	$\pm	0.05	$	&	-17.88	$\pm	0.19	$	&	0.83	$\pm	0.01	$	&	9.14	\\
RXCJ0751-3	&	0.78	$\pm	0.03	$	&	-17.49	$\pm	0.22	$	&	2.23	$\pm	0.01	$	&	8.88	\\
RXCJ1206-1	&	0.60	$\pm	0.03	$	&	-16.17	$\pm	0.25	$	&	1.56	$\pm	0.07	$	&	8.52	\\
RXCJ1223-1	&	0.63	$\pm	0.08	$	&	-16.67	$\pm	0.13	$	&	1.11	$\pm	0.03	$	&	8.78	\\
RXCJ1715-1	&	0.87	$\pm	0.05	$	&	-18.22	$\pm	0.10	$	&	1.33	$\pm	0.01	$	&	9.67	\\
RXCJ1715-2	&	0.80	$\pm	0.04	$	&	-17.98	$\pm	0.15	$	&	1.27	$\pm	0.01	$	&	9.79	\\
RXCJ1715-3	&	0.82	$\pm	0.08	$	&	-18.15	$\pm	0.17	$	&	1.55	$\pm	0.04	$	&	9.57	\\
RXCJ2214-1	&	0.82	$\pm	0.07	$	&	-17.84	$\pm	0.11	$	&	1.44	$\pm	0.01	$	&	9.43	\\
RXCJ2214-2	&	0.74	$\pm	0.06	$	&	-17.88	$\pm	0.11	$	&	1.50	$\pm	0.02	$	&	9.55	\\
RXCJ2214-3	&	0.72	$\pm	0.04	$	&	-15.73	$\pm	0.23	$	&	0.91	$\pm	0.02	$	&	8.54	\\

		\hline
  \end{tabular}
 \end{table}

\section{Results}

\begin{figure}
	\subfigure{\includegraphics[width=\columnwidth]{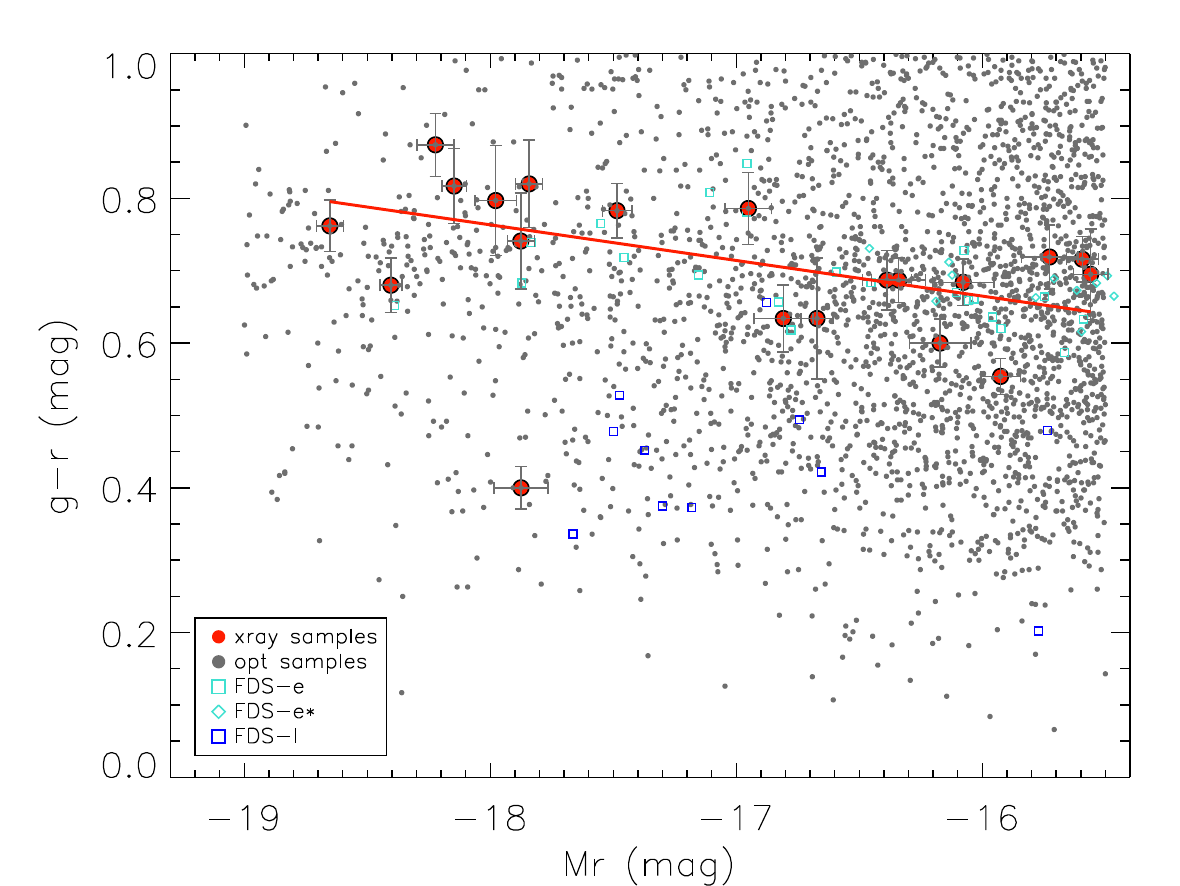}}
    \subfigure{\includegraphics[width=\columnwidth]{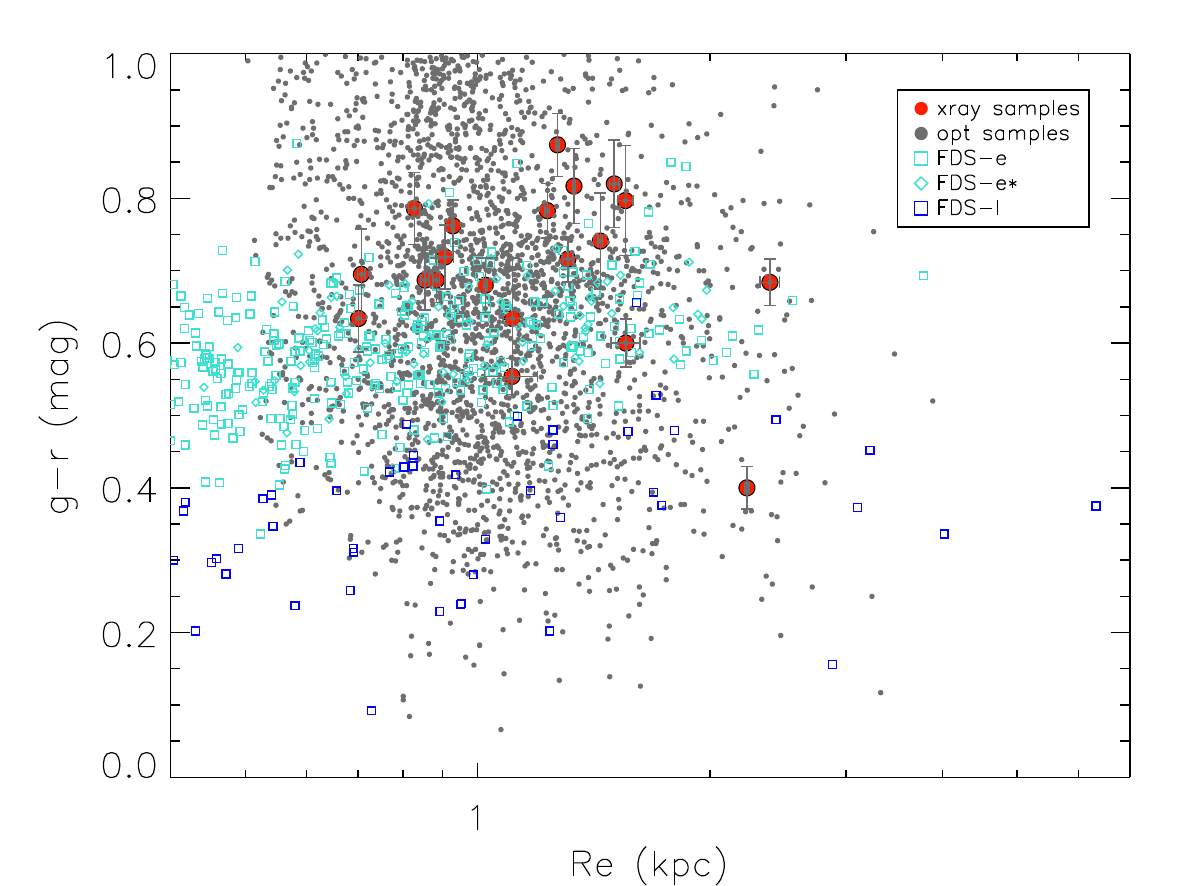}}
    \subfigure{\includegraphics[width=\columnwidth]{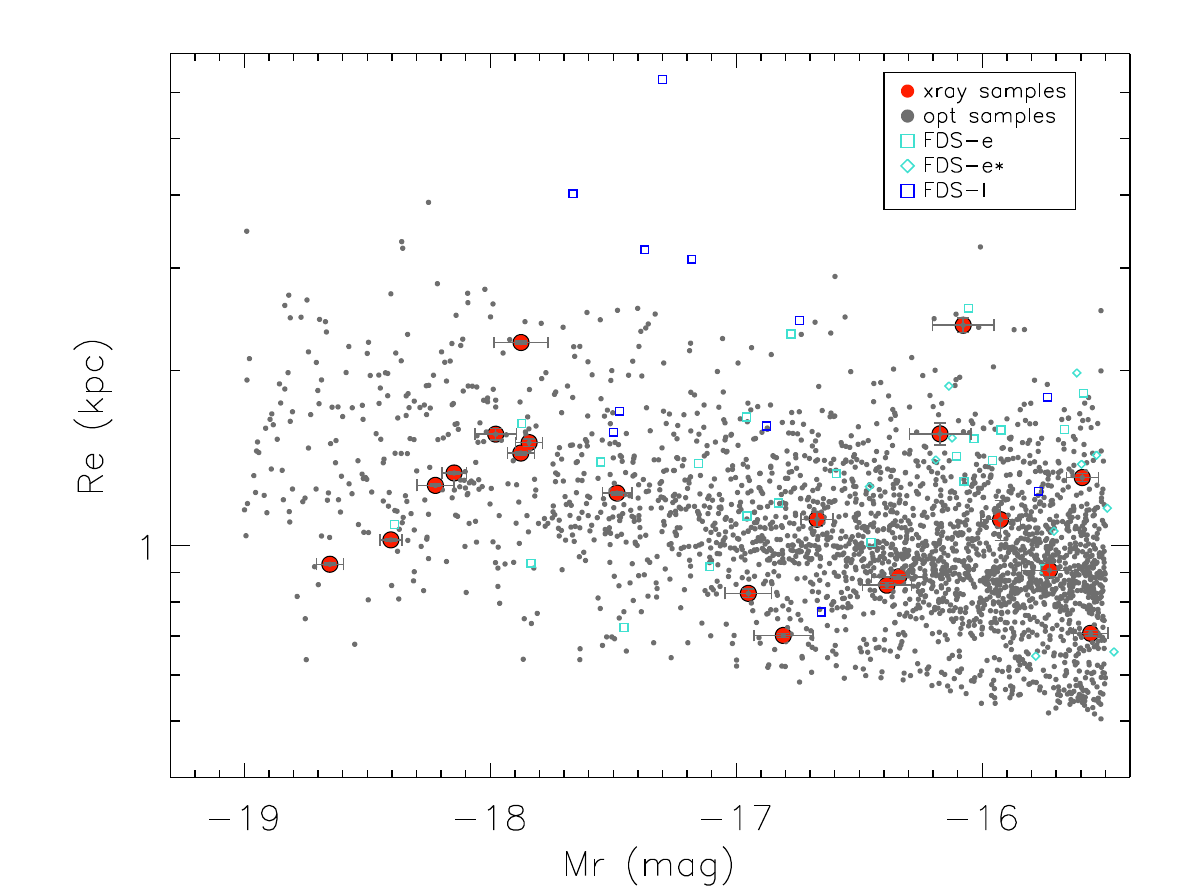}}

   \caption{Color$-$M$_r$, Color$-$\textit{R$_e$} and \textit{R$_e$}$-$M$_r$ diagram for the identified dwarf galaxies in all 12 clusters. Red points represent those with paired X-ray emission. Turquoise and blue symbols show the FDS early- and late-type dwarfs, respectively.}
    \label{fig:color_mag}
\end{figure}

We present the color-magnitude diagram (CMD) of all optically detected systems in Figure \ref{fig:color_mag}. We also indicate those 20 galaxies emitting X-rays on the same Figure. We find that X-ray emitting dwarfs cover this range of magnitude almost uniformly. On the other hand, the g$-$r colors of these dwarfs range in a rather narrow interval of 0.55 and 0.85. Note that the g$-$r color outlier with 0.40 (RXCJ0751$-$2), the estimated effective radius is one of the largest. Excluding this system, we observe a nearly linear trend between the g$-$r color and absolute magnitude. A linear fit to those 19 yields a slope of $-0.52\pm0.02$ which is the red sequence. We compare these relations with early- and late-type dwarf galaxies from the Fornax Deep Survey (FDS, \citealp{venhola.2018}). One clearly sees that early type dwarfs and our X-ray emitting samples form a red sequence, while the late type dwarfs are situated in a blue cloud below it. Our outlier dwarf galaxy is located within the region occupied by FDS late-type dwarf galaxies in Figure \ref{fig:color_mag}.

\begin{figure}
	\subfigure{\includegraphics[width=\columnwidth]{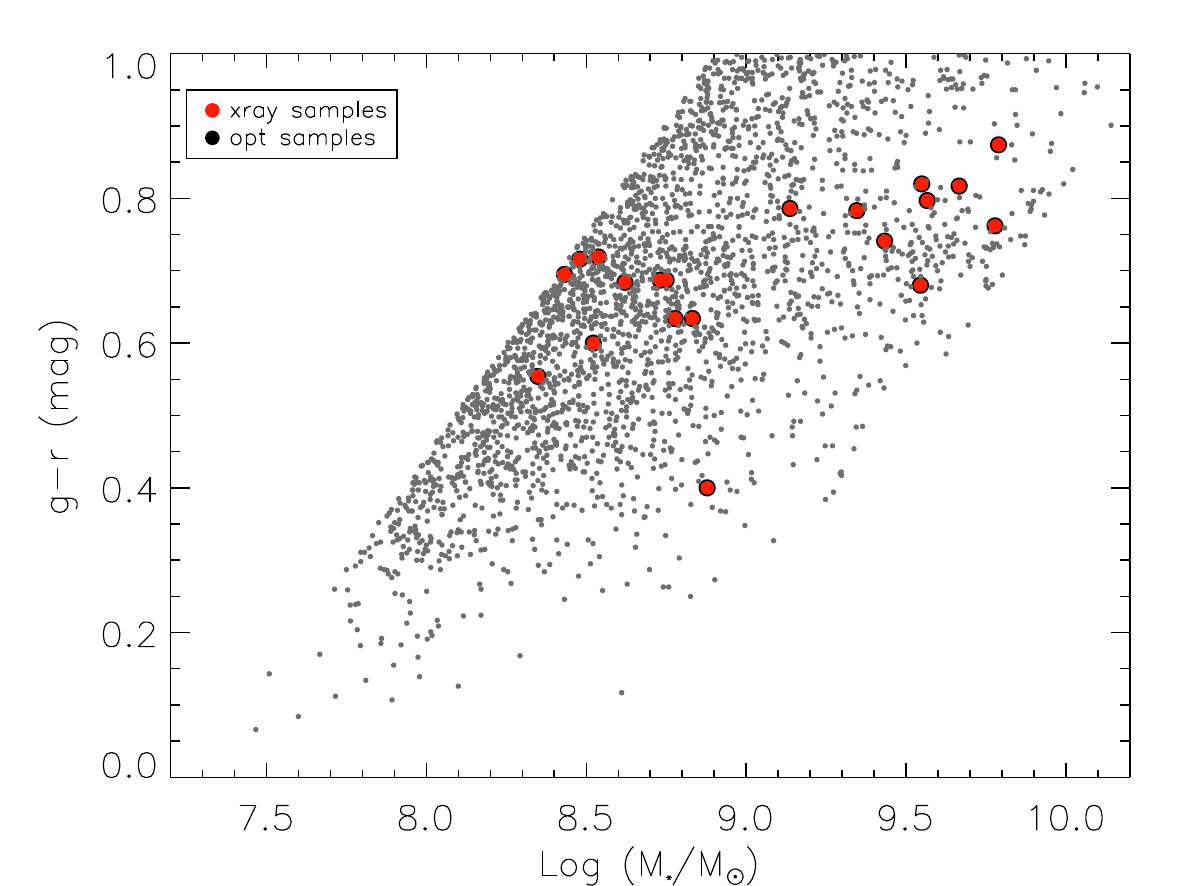}}
    \subfigure{\includegraphics[width=\columnwidth]{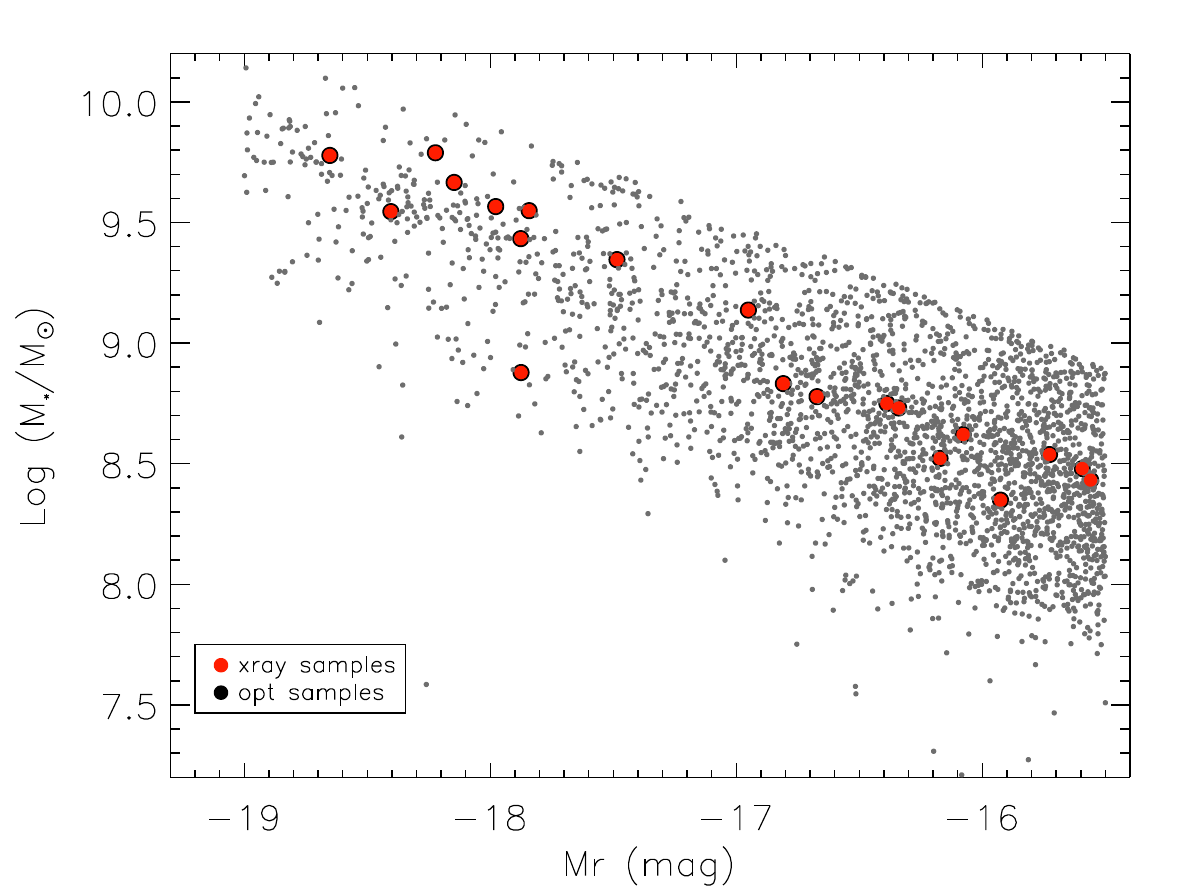}}
    \subfigure{\includegraphics[width=\columnwidth]{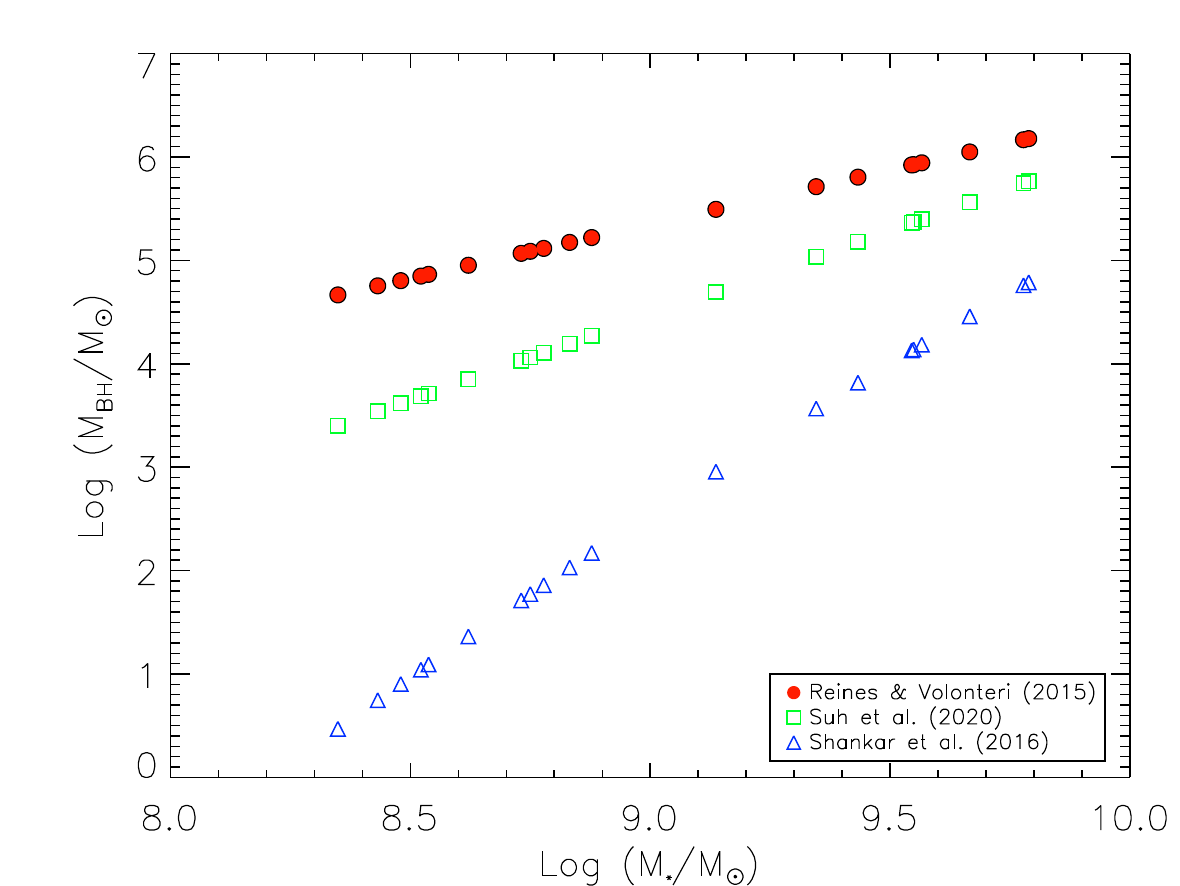}}

   \caption{Color$-$stellar mass \textit{(top)} and stellar mass$-$Mr diagrams \textit{(middle)} for the identified dwarf galaxies in all 12 clusters. Red points represent those with paired X-ray emission. Black hole mass as a function of stellar mass \textit{(bottom)} with three different assumptions. }
    \label{fig:color_mass}
\end{figure}

~

We present the color vs. effective radius behavior of the optically identified dwarf galaxies in Figure~\ref{fig:color_mag} (middle). We find that 80\% of the objects detected in X-ray have a size smaller than 1.5 kpc. They appear to be also clustered in $R_{e}$ $\sim$ 1.2 kpc, that is in general agreement with the size distribution of dwarf. Also in Figure \ref{fig:color_mag} (bottom), we show effective radius against absolute magnitudes of galaxies. It is noteworthy that the clustering of dwarfs towards smaller $R_{e}$ and high M$_{r}$ is not followed by those dwarfs identified in X-rays, indicating that the detected objects have a range in surface brightness (or density).

We plot stellar mass as a function of color, M$_{r}$ in the top and middle panels of Figure \ref{fig:color_mass}, respectively. Our X-ray emitting dwarf samples mass range from $\sim$ 2 $\times10^{8}$ $M_{\sun}$ to 6 $\times10^{9}$ $M_{\sun}$. We find no X-ray emitting dwarf galaxies below the galaxy mass of $M_{*}$ $\sim 10^{8.3}M_{\sun}$.

\begin{figure}
	\includegraphics[width=0.99\columnwidth]{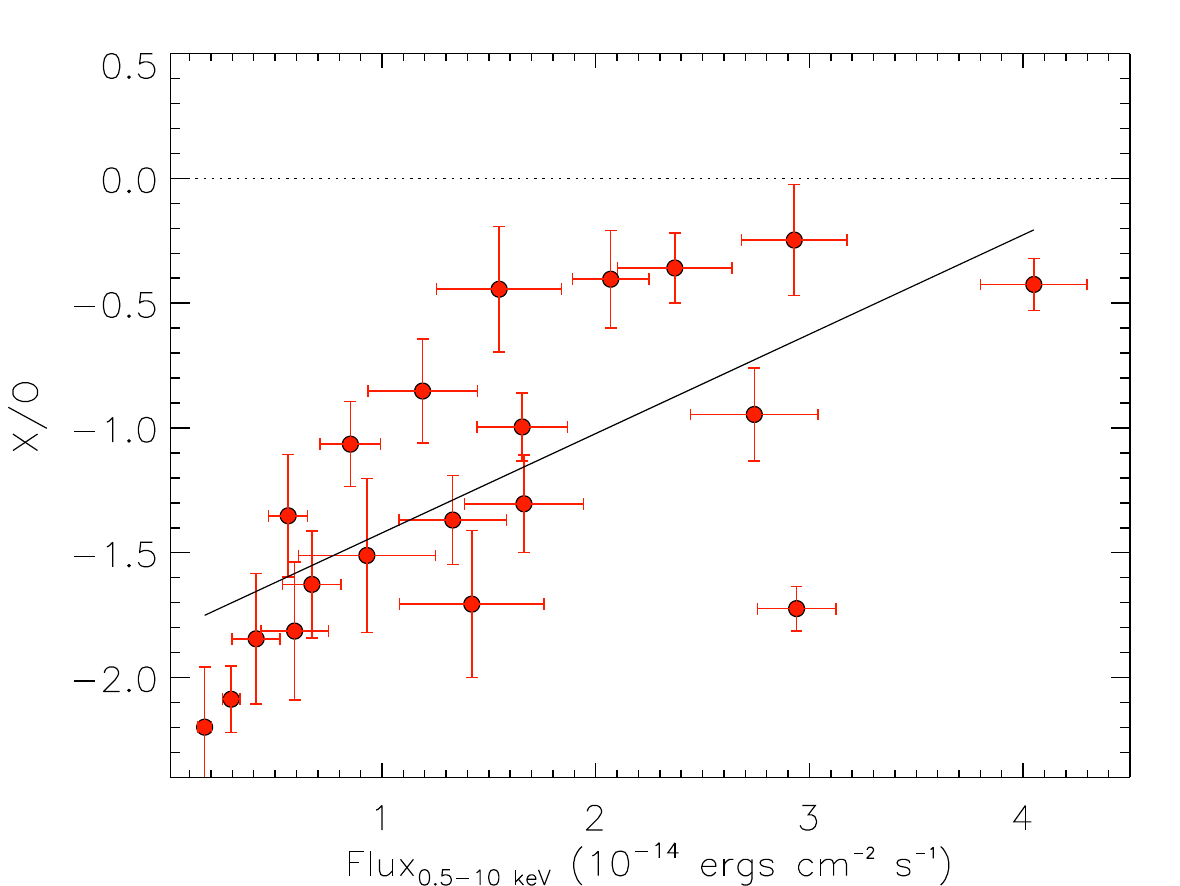}
   \caption{Plot of X-ray to optical flux ratio vs. the X-ray flux in the 0.5-10 keV band. The solid line is the best fit linear trend to these data points. The horizontal dotted line marks the level of equal X-ray and optical fluxes.}
    \label{fig:x_o_flux}
\end{figure}

We also studied the X-ray to optical emission (X/O) ratio as introduced by \cite{maccacaro.1988} as

    \begin{equation}
        X/O = \log(f_{X} / f_{opt}) = \log(f_{X}) + C + m_{r} / 2.5
    \end{equation}
where, \textit{f$_X$} is the X-ray flux in given energy band in ergs cm$^{-2}$ s$^{-1}$, \textit{m$_{opt}$} is the magnitude at the chosen optical wavelength and C\footnote{It is taken as 5.57 for r band from \cite{Haggard.2010}.} is a constant which depends on the specific filter used in the optical observations. We present X/O values of these 20 systems against their X-ray flux in the 0.5$-$10 keV band in Figure \ref{fig:x_o_flux}. We find that all X/O ratios in our sample are less than $-0.25$. Moreover, there is a positive correlation between X/O values and flux with Spearman's rank order correlation coefficient, r of 0.70 and chance occurrence probability, P of 4.5$\times10^{-4}$. A linear trend fit to these data points yields a slope of 0.69$\pm$0.19. We also investigated the trends of X/O and O/X with their corresponding optical flux values. We find that both ratios vary significantly with increasing optical flux as well. Therefore, X/O vs. X-ray flux is an indicative probe for both X-ray and optical emission.

An important advantage of BH-galaxy scaling relationships is their ability to estimate the mass of a black hole from readily measurable galaxy properties. This feature makes them a valuable tool for investigating BH properties and their role in galaxy evolution. In our case, it is, however, not possible to estimate the mass of a black hole (M$_{BH}$) using the relationship between M$_{BH}$ and stellar velocity dispersion ($\sigma_{*}$) (e.g., \citealt{ferrarese.2000,gultekin.2009,beifiori.2012}) due to the lack of available $\sigma_{*}$ measurements in our samples. Previous studies have shown a strong positive correlation between M$_{BH}$ and M$_{*}$ \citep{reines.2015,Shankar.2016,Suh.2020}. Here, we evaluate the relationship between M$_{BH}$ and M$_{*}$ with three different assumptions. \citet{reines.2015} suggested a relation between the masses of the black holes and their host galaxies (see their equation 5) using a sample of nearby inactive early-type galaxies and local AGN. They report that their M$_{BH}$ measurements have errors of $\sim$0.5 dex. \citet{Shankar.2016} used a combination of dynamical modeling and the virial method to calculate black hole masses (see their equation 6).  \citet{Suh.2020} investigate the relationship between black hole mass and galaxy total stellar mass up to a redshift z $\sim$ 2.5 for 100 X-ray-selected AGN sample and provides a relation via their equation 2.

\begin{figure}
	\includegraphics[width=0.99\columnwidth]{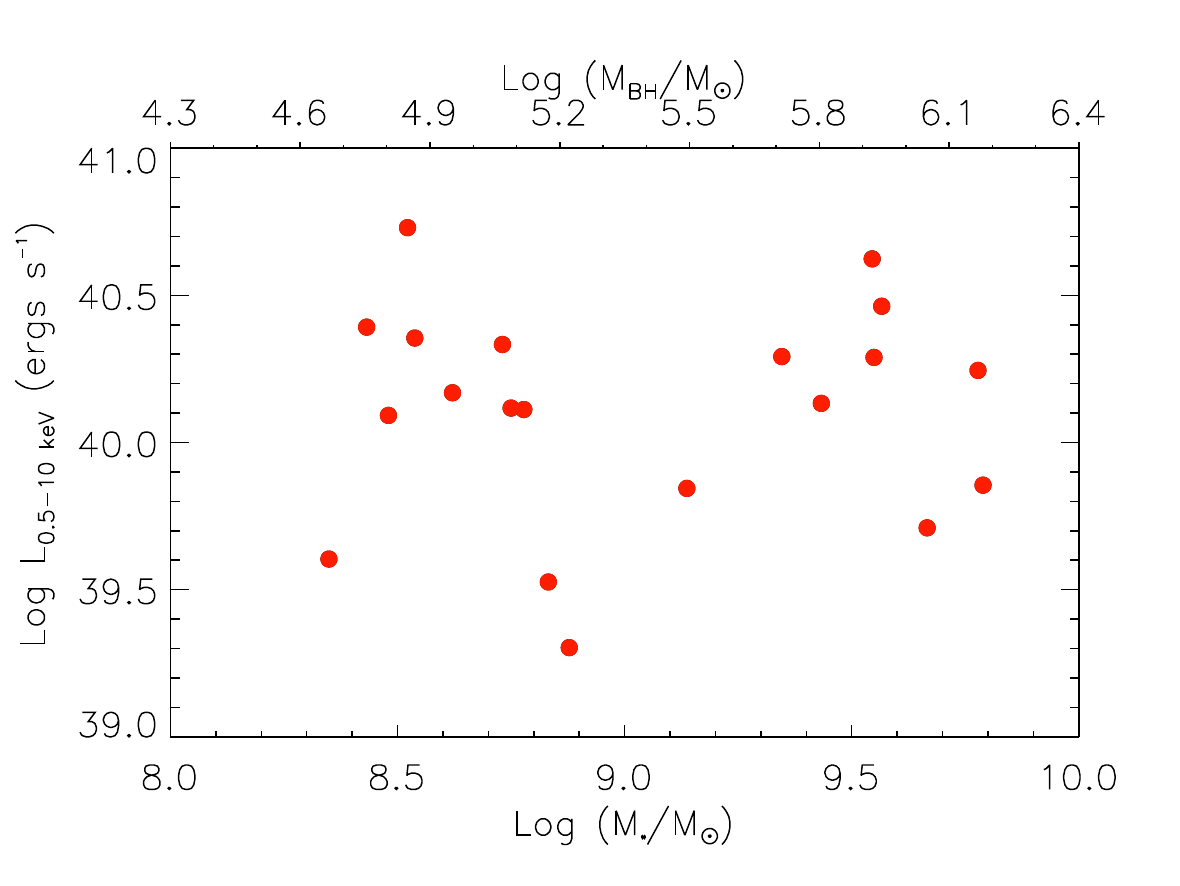}
   \caption{Plot of X-ray luminosity in the 0.5$-$10 keV band against the estimated mass of corresponding dwarf galaxies (lower scale), as well as black hole mass (upper scale).}
    \label{fig:lum_bhmass}
\end{figure}

Note that their sample of galaxies has total stellar masses of $10^{11-12}M_{\sun}$.
We present $M_{BH}$ estimates within these three approaches for the sample of 20 X-ray emitting galaxies in the bottom panel of \ref{fig:color_mass}. It is important to note that the
$M_{BH}$ vs. $M_{*}$ relation for dwarf galaxies is not established well. Nevertheless, the sample of \citet{reines.2015} is the only one that includes dwarf galaxies and extends to lower galaxy masses. Therefore, we employed the relation of \citet{reines.2015} and obtain an average black-hole-mass of 2.4$\times10^{5} M_{\sun}$ for our 20 galaxy sources (see Figure~\ref{fig:lum_bhmass}).

\section{Discussion and Conclusions}

Through our investigation of the correlation between optical and X-ray data of a sample of nearly galaxy clusters, we identified 2720 dwarf galaxies, twenty of which were found to emit also in X-rays. 
Earlier multiband investigations revealed by \citet{pardo.2016} revealed X-ray emission from ten dwarf galaxies at z < 1 out of 605 in the optical. They estimate an AGN fraction of $\sim$ 1\% for their sample of dwarf galaxies. They identified AGN candidates based on their hardness ratios and a hard X-ray photon index in the energy range between 2 and 10 keV. \citet{reines.2013} studied a sample of 151 dwarf galaxies (mass range from $10^{8.5}$ to $10^{9.5}$ M$\sun$, z < 0.055) with narrow and/or broad emission line signatures indicating the presence of an AGN and identified 10 dwarf galaxies with both narrow and broad emission line AGN signatures.  \citet{lemons.2015} cross-matched the \citet{reines.2013} parent sample with the Chandra Source Catalog and found 8 systems with nuclear hard X-ray emission at levels higher than expected from low-mass and high-mass X-ray binaries.  \citet{miller.2015} is another study that used a combination of optical and X-ray data to identify AGN in dwarf galaxies. They conducted a study on approximately 200 early-type dwarf galaxies in the local universe. These galaxies were selected optically using Hubble Space Telescope imaging as part of the AGN Multi-wavelength Survey of Early-type galaxies (AMUSE) surveys (see \citealt{gallo.2008,miller.2012}, for details on these surveys). They identified a total 23 AGN in their samples with mass range from about $10^{7}$ to $10^{9}$ M$\sun$. Recently, \citet{chen.2017} identified ten AGNs in low-mass galaxies from the NuSTAR serendipitous survey \citep{lansbury.2017}, which is capable of probing hard X-ray emission up to 24 keV.

The X-ray flux of our sample of 20 galaxies range from 1.7$\times10^{-15}$ erg cm$^{-2}$ s$^{-1}$ to 4.1$\times10^{-14}$ erg cm$^{-2}$ s$^{-1}$. The corresponding X-ray luminosities (considering isotropic emission at their corresponding distances) vary from 2$\times10^{39}$ erg s$^{-1}$ to 54$\times10^{40}$ erg s$^{-1}$. Note that even the lowest X-ray luminosity exceeds the Eddington luminosity limit for a typical neutron star (mass of 1.4 M$\sun$ and radius of 10 km).  
\cite{mezcua.2016} also analyzed star-forming dwarf galaxies in the COSMOS field up to z = 1.5, and found an excess of X-ray emission that is attributed to a population of accreting BHs, accounting for expected contributions from low-mass and high-mass X-ray binaries (LMXBs and HMXBs, respectively) as well as X-ray emission from hot gas.  

X-ray binaries (XRBs) are expected to make a significant contribution, primarily through the presence of young, HMXBs and and the latter by longer-lived low-mass XRBs. This is because LMXBs are typically found in early-type galaxies with very low star formation rates and ages exceeding a few Gyr (see review by \citealt{fabbiano.2006}). The luminosity of LMXBs can range up to $10^{38}$ erg s$^{-1}$, while HMXBs can have luminosities higher than $\sim10^{39}$ erg s$^{-1}$. We can consider the contribution of HMXBs according to our samples luminosity range. On the other hand, we expect that galaxies with larger SFR will have more significant contributions from HMXBs but dwarf galaxies are mostly passive and old systems. 

 \cite{Lehmer.2016} reported an intriguing observation that a larger population of HMXBs is found in regions with lower metallicity and the luminosity function of HMXBs at lower luminosity (L$_{x}$ < 10$^{38}$ erg s$^{-1}$) does not show significant sensitivity to changes in metallicity. With the purpose of determining the impact of metallicity, we calculate the gas-phase metallicity\footnote{Here, we adopt the term "metallicity" to refer to the gas-phase oxygen abundance, which is measured in units of 12 + log (O/H). The ratio of O/H represents the abundance of oxygen relative to hydrogen by number. The solar metallicity is defined in this scale as 8.69 (Allende Prieto et al., 2001), while the metallicities of the Large Magellanic Cloud (LMC) and the Small Magellanic Cloud (SMC) are 8.4 and 8.0, respectively (Garnett, 1999).} for three out of the total of twenty galaxies in our study. This limited number of calculations was due to the lack of optical spectral observations for the majority of the galaxies. The calculated gas-phase metallicity values for the three galaxies are as follows: 8.48 for A1367-1, 8.43 for RXCJ0751-1, and 8.47 for RXCJ2214-2. A detailed descriptions of analysis is given in Appendix~\ref{sec:appendix}.

The relation between size and color in dwarf galaxies in cluster environments is similar to that observed in the general population of dwarf galaxies. According to the analysis conducted, the dwarf galaxies in the cluster environment are on the red sequence, which means that they are relatively old and passive objects. It does not appear that the contribution of HMXBs plays a significant role in our sample.

In order to investigate the effect of AGN, we calculated the black hole masses of galaxies using three approaches in the literature.

The virial method has been a useful tool in estimating the masses of these objects in quiescent galaxies. \citet{reines.2015} employed this method to estimate black hole masses and found a linear $M_{BH}$--$M_*$ relation, indicating that the black hole mass scales proportionally with the host galaxy's stellar mass. However, they did not take into account the bias introduced by resolving the black hole's sphere of influence. \citet{Shankar.2016} adopted a combination of dynamical modeling and the virial method to estimate black hole masses in galaxies. Moreover, they accounted for the bias introduced by resolving the black hole's sphere of influence, resulting in a lower normalization than that derived by \citet{reines.2015}. Note that the highest stellar mass in our sample is 1.4$\times10^{10}$ M$\sun$. However, the de-biased $M_{BH}$--$M_*$ relation by \citet{Shankar.2016} was derived for galaxies with $M_*$ > 2$\times10^{10}$ M$\sun$. This relation yields significantly lower mass for  the low mass systems in our sample, down to stellar mass black hole regime. To further investigate the $M_{BH}$--$M_*$ relation, \citet{Suh.2020} employed a Bayesian approach to estimate black hole masses in early-type galaxies up to z $\sim$ 2.5. Considering the stellar mass range and redshifts, the scaling relation by \citet{reines.2015} is the best resemblance to our sample among the three approaches. In this framework, the highest $M_{BH}$ reaches the level of 10$^{6.2}M\sun$. Note that even higher black hole masses were reported in dwarf galaxies using optical diagnostics \citep{reines.2013}, X-ray observations \citep{mezcua.2018,Mezcua.2023} and radio observations \citep{reines.2020}.

The X-ray to optical luminosity ratio is a useful tool for characterizing the properties of dwarf galaxies, and there have been several studies in the literature that have explored this ratio for these systems. X/O can provide information about the source of the X-ray emission, such as hot gas, star formation, or an AGN. In general, dwarf galaxies exhibiting high X-ray to optical emission ratios are presumed to contain abundant amounts of hot gas, which is likely due to a plethora of phenomena such as supernova explosions, tidal interactions, or other factors. Dwarf galaxies with low X-ray to optical emission ratios, on the other hand, may contain little hot gas and may be dominated by star formation or AGN activity \citep{jeltema.2005,mineo.2012}. Nonetheless, it's worth mentioning that the X-ray to optical emission ratio can fluctuate extensively between diverse dwarf galaxies and can also be affected by a range of parameters, such as the star formation history, the presence of an AGN, and the environment. For example, dwarf galaxies in dense cluster environments may have lower X-ray to optical emission ratios due to ram pressure stripping, which removes hot gas from these systems \citep{liu.2019}.

We find that the dwarf galaxies in our sample have X/O ratios less than -0.25 and there exists a positive correlation between the X/O values and the flux of these galaxies. X-rays in galaxy scales are usually produced through accretion processes onto the central black holes. An increased mass accretion rate would yield an increase in X-rays (from the inner portions of the disk), as well as and increase in optical (from the outer accretion disk). Therefore, the positive correlation between X/O values and X-ray flux levels observed could be due to mass transfer rate onto the central compact object.

The use of multiwavelength observations, especially targeting the optical and X-ray wavelength ranges, has emerged as a potential probe for elucidating the intrinsic characteristics of X-ray sources. Notably, Ultra-Luminous X-ray Sources (ULXs) display X/O ratios primarily spanning the range of 1.5 to 2.5 indicating that the X-ray flux is much higher than the optical flux in these objects \citep{Feng.2008}. With the resulting values of our X-ray–to–optical flux ratios, we can conclude that our X-ray sources do not indicate the characteristics typically associated with ULXs. On the other hand, an X-ray source with X-ray-to-optical flux ratios (considering the R-band as the reference for optical flux) ranging from -1.0 to 1.7 is characteristically identified as an AGN \citep{maccacaro.1988,Lehmer.2016}. 
Such a wide range of values suggests that AGN generally exhibits a broader distribution of X-ray to optical flux ratios, encompassing cases where the optical emission can exceed that of X-rays. Moreover, based on the position of our limited sample in the BPT diagram (\citealt{Baldwin.1981}; see Fig.~\ref{appendix:fig:bpt}), a commonly employed tool for classifying galaxies based on the primary sources of ionizing radiation, one would conclude that they would be classified as AGN rather than star-forming galaxies.

In order to achieve a more comprehensive understanding of the characteristics of dwarf galaxies, it is crucial to have extensive and detailed surveys that can provide a significant number of objects to study their kinematics, stellar populations, and metallicity. These surveys can enable the study of properties such as substructure within the dwarf population, which can obtain insights into how the cluster environment influences the formation and evolution of these galaxies and the reasons behind their X-ray emissions.

Further observations and studies are essential to better understand the nature of these X-ray emitting components. Nonetheless, hope is on the horizon as the imminent WEAVE nearby spectroscopic cluster survey holds the promise of providing us with a wealth of data. This survey will comprehensively cover all the clusters of our photometric survey and endow us with thousands of kinematic properties, such as velocity dispersion per cluster. With the help of this upcoming data, we could unravel the intricate properties of dwarf galaxies and gain invaluable insights into the role of their environment in shaping their evolution.

\section*{Acknowledgements}

We thank the anonymous referee for valuable comments and suggestions that improved the clarity and impact of our results. \c{S}. \c{S}. (Aydemir) acknowledges support from through 2218-National Postdoctoral Research Fellowship Program 
under project number 118C553 from the The Scientific and Technological Research Council of Turkey (T\"UB\.ITAK). A.A. acknowledges support from the ACIISI, Consejer\'{i}a de Econom\'{i}a, Conocimiento y Empleo del Gobierno de Canarias and the European Regional Development Fund (ERDF) under the grant with reference PROID2021010044. NCC  acknowledges support from  \textquotesingle Proyecto comit\'{e} mixto ESO - Gobierno de Chile\textquotesingle, N° 21119.

\section*{Data Availability}
Upon request to the corresponding authors, the optical data supporting this article will be made available. The original, unprocessed data is currently stored in the Isaac Newton Group  Archive. The X-ray data underlying this article are available in the Chandra Data Archive (https://cxc.harvard.edu/cda/) by searching the observation identifiers (ObsID) listed in Table\ref{tab:result_xray} in the Search and Retrieval interface, ChaSeR (https://cda.harvard.edu/chaser/).



\bibliographystyle{mnras}
\bibliography{mnras_template.bib} 




\appendix

\section{Measuring gas-phase metallicity}
\label{sec:appendix}
In order to characterise the origin of the X-ray fluxes we detected in dwarfs, we analyze their optical spectra.  Unfortunately, the availability of optical spectra has been limited for the sources in our study. We were constrained to employ the SDSS-DR10 spectroscopic main galaxy sample data set, which provided information for only six galaxies (A1367-1, RXCJ0751-1, RXCJ1715-1, RXCJ1715-2, RXCJ2214-1, and RXCJ2214-2). To measure the emission-line fluxes, we utilize the penalized pixel-fitting (pPXF) \citep{Cappelari.2004} method and fit the observed spectra using a combination of templates from the E-MILES stellar library \citep{vazdekis.2016}. These templates cover the wavelength range of 3525$-$7500 Angstrom with a constant spectral resolution of 2.50 Angstrom FWHM, similar to the SDSS spectral resolution. Among these six galaxies, only three (A1367-1, RXCJ0751-1, and RXCJ2214-2) had a good continuum fit thereby enabling us to calculate their emission lines and metallicity with reasonable accuracy.

\begin{figure}
	\includegraphics[width=0.99\columnwidth]{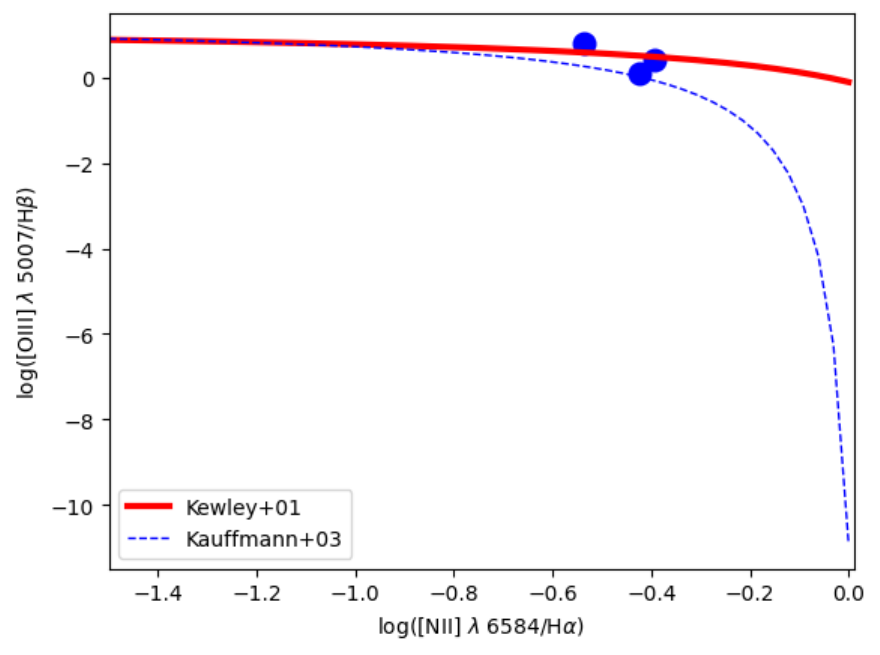}
   \caption{BPT diagram for our three X-ray detected dwarf galaxies in SDSS. The red solid line is the edge of the composite ionization region from Kewley et al. (2001). The blue dotted line shows the edge of the star formation photoionization region, derived by Kauffmann et al. (2003).}
    \label{appendix:fig:bpt}
\end{figure}

We employed standard emission-line ratios of [NII]$\lambda$6584/H$\alpha$ and [OIII]$\lambda$5007/H$\beta$. These ratios are commonly represented in line ratio diagrams known as "BPT diagrams" \citep{Baldwin.1981}. An elevated value of [NII]/H$\alpha$ (>0.6) is specifically for classifying the presence of AGN activity, while [OIII]$\lambda$5007/H$\beta$ helps distinguish between low-ionization nuclear emission-line regions (LINERs; Heckman 1980 and Seyfert-type objects, which LINERs have [OIII]$\lambda$5007/H$\beta$ < 3). \citet{Kauffmann.2003} proposed two demarcation lines (see Fig.\ref{appendix:fig:bpt}) on the BPT diagram to separate AGNs, star-forming galaxies, and AGN-star-forming composites. The upper line represents the theoretical upper limit for starburst galaxy emission-line ratios based on \citet{Kewley.2001}. Galaxies positioned above this line in the BPT diagram are expected to exhibit a significant contribution from an AGN to their emission-line flux. The second line, which corresponds to typical star-forming galaxies, is a modified version of the \citet{Kewley.2001} line proposed by (\citealp{Kauffmann.2003}). The optical classification of galaxies lying between these two lines is called the AGN-starburst composite. Composite galaxies are anticipated to have approximately 41\% of their [OIII]$\lambda$5007 luminosity density and 11 \% of their H$\alpha$ luminosity density attributed to an AGN \citep{Brinchmann.2004}.

To estimate the gas-phase metallicity for three galaxies out of six, the location of those galaxies on the BPT diagram based on \citealp{Kauffmann.2003} are in the AGN part. In this case, we consider the AGN gas-phase metallicity relation based on strong emission lines ([OIII]$\lambda$5007, H$\alpha$, H$\beta$, and [NII]$\lambda$6584), according to \citep{Storchi-Bergmann_1998} and also we consider the correction in the gas-phase metallicity derivation according to the electron density effects on the mentioned calibration relation based on [SII]$\lambda$6716 and [SII]$\lambda$6731 ratio. We presented the BPT diagrams (see Fig.~\ref{appendix:fig:bpt}), emphasizing the reliance on the BPT diagnostic diagram for gas-phase metallicity estimation. We also reported the strong gas-phase metallicity values for these galaxies.

For the remaining galaxies, however, difficulties were encountered in estimating gas-phase metallicity due to the lack of detectable strong emission lines. We tested the different approaches such as multiple component fits, high-degree polynomial fits, and boundary constraints on both stellar velocity and gas velocity. Nonetheless, none of these methods made the emission lines detectable. Moreover, we acknowledged the challenges associated with detecting certain emission lines ([OIII]$\lambda$4363 or [OII]$\lambda$$\lambda$7320,7330) for the metallicity estimation. Applying the \citet{Kauffmann.2003} criterion would classify all galaxies as AGNs, while \citet{Kewley.2001} would classify two galaxies as SF and one as an AGN. The gas-phase metallicity values are 8.48, 8.43, and 8.47 for A1367-1, RXCJ0751-1, and RXCJ2214-2, respectively based on AGNs metallicity calibration relation. To be cautious enough, if we consider the combination of two fits on the BPT diagram, there is no straightforward gas-phase metallicity estimation in the AGN-starburst composite region of the BPT diagram. Moreover, we estimated the gas-phase metallicity based on the metallicity calibration of star-forming galaxies \citep{curti.2016} for two galaxies, while considering the fit on the BPT diagram established by \citet{Kewley.2001}. The results including errors of 0.1 remained in sub solar region.

\bsp	
\label{lastpage}
\end{document}